\documentclass{iau}
\usepackage{graphicx}
 \usepackage{epstopdf}

\title[HD\,50230 --- a failure of richness] 
{Interpretation of the oscillation spectrum of HD 50230 --- a failure of richness}

\author[Szewczuk, Daszy\'nska-Daszkiewicz \& Dziembowski]   
{Wojciech Szewczuk$^1$, Jadwiga Daszy\'nska-Daszkiewicz$^1$\\
 \and Wojciech Dziembowski$^2$}

\affiliation{$^1$Instytut Astronomiczny, Uniwersytet Wroc{\l}awski, Wroc{\l}aw, Poland
\\ email: {\tt szewczuk@astro.uni.wroc.pl, daszynska@astro.uni.wroc.pl} \\[\affilskip]
$^2$Warsaw University Observatory, Warsaw, Poland \\email: {\tt wd@astrouw.edu.pl}}

\pubyear{2014}
\volume{301}  
\pagerange{1--4}
\jname{Precision Asteroseismology}
\editors{J.A. Guzik, W.J. Chaplin, G. Handler \& A. Pigulski, eds.}
\begin{document}

\maketitle

\begin{abstract}
Attempts to interpret the observed oscillation spectrum of the SPB
star HD\,50230 are reported. We argue that a nearly equidistant period
spacing found in the oscillation spectrum of the star is most likely
accidental. The observed period distribution requires excitation of modes
with the degree $\ell>4$.
Much more may be learned from the rich oscillation spectrum of the star
but most of the work is still ahead of us.
\end{abstract}

\firstsection 

\section{Introduction}
HD\,50230 is a star of the B3\,V spectral type and a visual brightness of
8.95 mag. The star had been regarded as constant until its first satellite
observations were conducted. \cite[Degroote et al.~(2010,
2012)]{Deg2010} detected more than 500 significant peaks in the \textit{CoRoT} data.
With their spectroscopic observations they discovered that
HD\,50230 is a double-lined spectroscopic binary with projected
equatorial velocity of 7 and 117~km\,s$^{-1}$ for the primary and
secondary, respectively. For the primary component, they
determined effective temperature of $T_\mathrm{eff,1}=18000\pm 1500$~K
and surface gravity of $\log g_1=3.8\pm 0.3$~dex. For the secondary component
they found only an upper limit $T_\mathrm{eff,2} \le 16000$~K and assumed $\log g_2\approx 4$~dex.

\section{Peaks almost uniformly spaced in period}
In the rich oscillation spectrum of HD\,50230 \cite{Deg2010} extracted
eight peaks almost uniformly spaced in period. Invoking the
asymptotic theory, they interpreted these peaks as a
sequence of modes with the same spherical harmonic degree, $\ell$,
azimuthal order, $m$, and consecutive radial orders, $n$. Assuming
$\ell=1$ and $m=0$ they found that main-sequence star models with a mass of 7--8~$M_\odot$
can reproduce the observed period spacing.

To test Degroote's interpretation we re-analysed the \textit{CoRoT} data.
Using the Lomb-Scargle periodogram we found 515 significant frequency
peaks comparing to 556 frequencies found by Degroote at al. (2012).
Most our frequencies are consistent with the Degroote's determinations.
Surprisingly, we found in our set many sequences of peaks
nearly uniformly spaced in periods. Unfortunately, these sequences do not
yield sufficient clues to mode identification and are likely accidental.

\begin{figure}
\begin{center}
 \includegraphics[angle=-90, width=\textwidth]{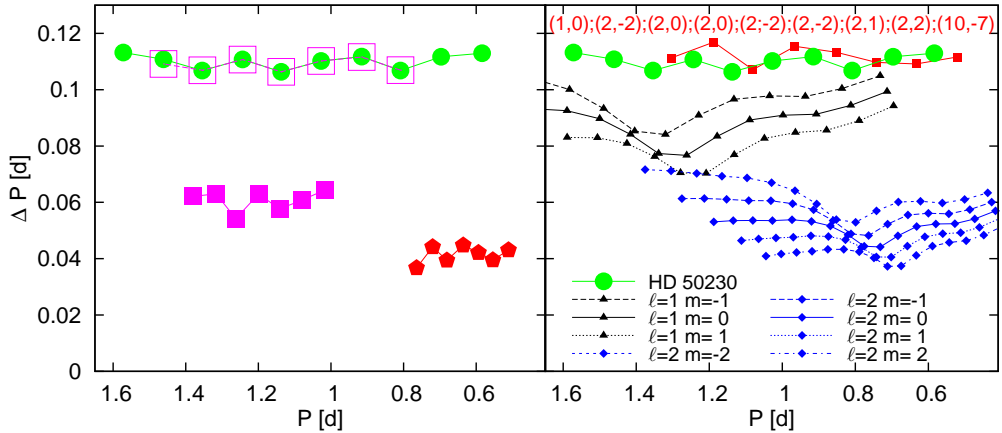}
 \caption{\textit{Left:} Nearly equidistant period spacings found in the \textit{CoRoT} photometric data of the star HD\,50230.
               Filled circles, squares and pentagons denote equidistant series found in the set of periods determined
               by us whereas the series published by \cite{Deg2012} is marked with open squares.
               \textit{Right:}
               Period spacings, in the model with a mass, $M=5.0$~$M_\odot$,
 effective temperature, $\log T_\mathrm{eff}=4.227$, luminosity, $\log L/L_\odot=2.77$,
 metallicity,  $Z=0.015$ and rotational velocity, $V_\mathrm{\rm rot}=10$ km s$^{-1}$.
Squares show selected modes with various degrees and azimuthal orders (numbers in brackets) with
spacings similar to that found in HD\,50230 (dots).
For comparison, sequences of consecutive dipolar (triangles) and quadruple (rhombs) modes are shown.
               }
   \label{fig1}
\end{center}
\end{figure}

In the left panel of Fig.\,\ref{fig1}, we present examples of three sequences (filled symbols).
For comparison, as open squares, a sequence found by \cite{Deg2012} is also shown.
We chose these sequences because if we assume that frequencies with the period spacing of $\Delta P\approx 0.11$~d
are dipole modes then, according to the asymptotic theory, those with $\Delta P\approx$ 0.06 and 0.04~d
correspond to modes with $\ell=2$ and 3, respectively. Unfortunately, they have some  common frequencies. Thus, at best,
there  must be some missing modes.

For further analysis, we selected nine frequencies from the sequence with $\Delta P\approx 0.11$~d.
Two frequencies with the shortest periods were omitted because they have too low radial orders and may not follow
the asymptotic theory. Next, we tried to reproduce these nine frequencies adopting two approaches.
Neglecting the effects of rotation, we reached the best fits for models with the following parameters:
$M =7.7$~$M_\odot$, $Z=0.030$, $\alpha_{\rm ov}=0.0$, $\log T_\mathrm{eff}=4.2337$ and
$M =7.1$~$M_\odot$, $Z=0.025$, $\alpha_{\rm ov}=0.2$, $\log T_\mathrm{eff}=4.2189$.
Here, $Z$ is the metal abundance by mass fraction and $\alpha_{\rm ov}$ is the overshooting parameter from the convective core
expressed in the terms of pressure scale height, $ H_p$. However, the corresponding $\chi^2\approx2\times10^4$ do not
allow us to accept these solutions.
A better fit was obtained when we took into account rotational splitting. Then, interpreting the sequence as
retrograde dipole modes for a model with
$M=5.4$~$M_\odot$, $Z=0.010$, $\alpha_{\rm ov}=0.6$, $\log T_\mathrm{eff}=4.2561$, and $V_\mathrm{rot}=25$ km s$^{-1}$
we got  $\chi^2\approx6\times10^3$. These large values of $\chi^2$ result from very small frequency errors.

\section{Theoretical modes nearly equally spaced in period}
Since oscillation spectra of high- and moderate-order g modes are dense,
we may expect that modes with different values of spherical harmonic degree $\ell$, and
azimuthal order $m$, can accidentally form period sequences nearly equally spaced in period.
To demonstrate such cases, we calculated pulsation modes with $\ell=1-10$ for models with $M=5$~$M_\odot$
and different rotational velocities. Then, in the same way as we did for the observations,
we searched for modes equally spaced in period, considering only unstable modes.
We found many sequences with different mean period spacings composed of modes
with different pairs ($\ell,m$). An example is shown in the right panel
of Fig.\,\ref{fig1} as squares. The nearly constant period spacing
is accidental and in no way is related to the asymptotic property of
g modes. The calculated sequences at fixed $\ell$ and $m$,
depicted in the same figure, show lower mean spacing and  much larger
deviation from constancy. These results support the conclusion that, in
dense oscillation spectra, appearance of equally spaced sequences may
be accidental and they should be treated with caution.

\section{Comparing histograms of observed and calculated frequencies}

\begin{figure}
  \begin{center}
   \includegraphics[angle=-90, width=\textwidth]{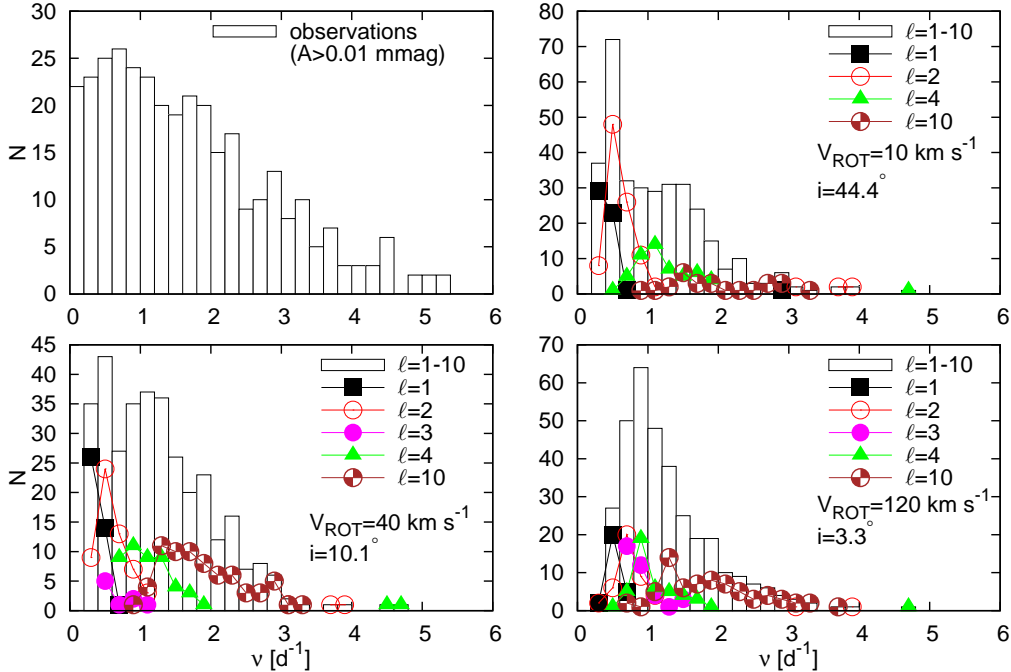}
  \caption{A comparison of histograms for frequencies detected in HD\,50230
(upper left) and frequencies of unstable modes in the model with
$M=6.95 M_\odot$, $\log T_\mathrm{eff}=4.255$, $Z=0.015$ and $\alpha_{\rm ov}=0$ and three
rotational velocities: 10 km s$^{-1}$, 40 km s$^{-1}$ and 120 km s$^{-1}$. In the theoretical histograms,
N stands for the number of all unstable modes with $\ell$ from 1 to 10
and $A\ge0.01$ mmag. The values of the inclination angle, $i$, result from $V_{\rm rot}\sin i = 7$ km s$^{-1}$.}
\label{fig3}
\end{center}
\end{figure}

When we realized that the observed period spacing cannot be interpreted
according to the asymptotic theory we lost the clue to mode identification
which is prerequisite for deriving seismic constraints on stellar
parameters. Since the prospects for progress looked to us rather
grim, we decided to look for a different application of the rich
frequency data on HD\,50230.

One unsolved problem in the stellar
pulsation theory is the amplitude limitation in stars with a large
number of unstable modes. The observed distribution of peaks over
period ranges may yield an important hint
leading us to the solution of this difficult problem. To this aim
the observed distribution must be confronted with simulations based
on linear nonadiabatic calculations for models constrained by
measured values of $\log T_\mathrm{eff}$, $\log g$ and
$V_{\rm rot}\sin i$. The linear calculations do not yield mode
amplitudes and this information we want to extract from data. We
should search for the best fit to the observed distribution assuming
various mode-selection principles. The set of considered modes,
which all must be unstable, may be terminated at some sufficiently
high $\ell$ if the observational threshold cannot be reached.

In the histograms shown in Fig.\,2, we assume the random distribution of
the r.m.s. amplitude of relative variations of the stellar radius, {\it the intrinsic amplitude of a mode},
and considered modes with the photometric amplitudes $A\ge0.01$ mmag.
We included modes with degrees up to $\ell=10$, because our simulations showed
that for higher $\ell$, the amplitudes in the \textit{CoRoT} band do not exceed the value of $A=0.01$ mmag.
We use the same model reproducing the central values of $\log T_\mathrm{eff}$ and $\log g$
(see Sect.\,1) for the primary but with three distinct rotation rates: $V_{\rm rot}=10,~40,~120$ km s$^{-1}$.
The values of the inclination angle result from $V_{\rm rot}\sin i = 7$ km s$^{-1}$ which was kept constant.
These rotation rates are higher than those estimated by \cite[Degroote et al.~ (2010, 2012)]{Deg2010}
based on analysis of high-order p modes. However, their estimate is uncertain
and refers to the different part of the interior.
The effects of rotation were included in the framework of
the traditional approximation, e.g., \cite{Lee1997}, \cite[Townsend (2003, 2005)]{Tow2003, Tow2005}.
For comparison, the observed histogram was shown in the left upper panel of Fig.\,2.

As one can see, we were unable to reproduce the observed distribution of frequencies.
Independently of rotation, we have a shortage of unstable and ``visible'' ($A\ge0.01$ mmag) theoretical modes with the shortest
as well as with highest frequencies, above about 2 d$^{-1}$. For higher rotation rate, one could expect
that retrograde modes complement the lowest frequency range. But the higher values of $V_{\rm rot}$ imply the lower values of the inclination angle
which in turn favor the axisymmetric modes ($m=0$). We considered also a possibility that some
of the low-amplitude peaks may result from combinations and found it unlikely.
Our results give certain limits on the intrinsic amplitudes of modes.

\section{Conclusions}
In dense oscillation spectra such as in the case of HD 50230, equidistant period spacings
can be very likely accidental and one should be cautious when interpreting such structures.
Although the dream of rich oscillation spectra in the B-type pulsators has come true, we still
do not have any clue to identify angular numbers of observed frequencies.
Without additional observations which would allow for mode identification, a reliable seismic stellar model
of the star cannot be constructed.

We see prospects for gaining new insight into nonlinear mode
selection in stars from available data on HD 50230. Our efforts
toward explaining the observed distribution of peaks in its
oscillation spectrum will continue.

\begin{acknowledgments}
WD was supported by Polish NCN grant DEC-2012/05/B/ST9/03932.
WS was supported by Polish NCN grant 2012/05/N/ST9/03905.
Calculations have been carried out using resources provided by Wroc{\l}aw Centre for Networking and Supercomputing (http://wcss.pl), grant No.~265.
\end{acknowledgments}

\end{document}